\documentclass[aps,prl,twocolumn,superscriptaddress,showpacs]{revtex4-1}
\usepackage{amsmath}
\usepackage{color}
\usepackage{graphicx}
\usepackage{epstopdf}
\usepackage{epsfig}
\usepackage{amsfonts}
\usepackage[naturalnames]{hyperref}
\usepackage{hypcap}
\usepackage{verbatim}

\makeatletter
\newcommand{\colorcaption}[2][]{%
  \begingroup%
  \renewcommand{\@caption@fignum@sep}{ (color online). }%
  \caption[#1]{#2}%
  \endgroup%
}
\makeatother

\begin{document}

\title{Dynamical Detection of Topological Phase Transitions in Short-Lived Atomic Systems}

\author{F.~Setiawan}
\email{setiawan@umd.edu}
\affiliation{Department of Physics, Condensed Matter Theory Center and Joint Quantum
Institute, University of Maryland, College
Park, Maryland 20742, USA}
\author{K.~Sengupta}
\affiliation{Theoretical Physics Department, Indian Association for
the Cultivation of Science, Jadavpur, Kolkata-700032, India}
\author{I.~B.~Spielman}
\affiliation{Joint Quantum Institute, National Institute of
Standards and Technology, and University of Maryland, Gaithersburg,
Maryland, 20899, USA}
\author{Jay D.~Sau}
\affiliation{Department of Physics, Condensed Matter Theory Center and Joint Quantum
Institute, University of Maryland, College
Park, Maryland 20742, USA}
\date{\today}

\begin{abstract}
We demonstrate that dynamical probes provide direct means 
of detecting the topological phase transition (TPT) between conventional 
and topological phases, which would otherwise be difficult to access 
because of loss or heating processes. We propose to avoid such heating 
 by rapidly quenching in and out of the short-lived topological phase across the transition 
that supports gapless excitations. 
Following the quench, the distribution of excitations in the final conventional phase
 carries signatures of the TPT. 
We apply this strategy to study the TPT into a Majorana-carrying topological phase predicted in one-dimensional
spin-orbit-coupled Fermi gases with attractive interactions.
The resulting spin-resolved momentum distribution, computed by self-consistently solving the time-dependent
Bogoliubov--de Gennes equations, exhibits Kibble-Zurek scaling and 
St\"{u}ckelberg oscillations characteristic of the TPT. We discuss parameter regimes where the TPT is experimentally accessible.

\end{abstract}


\maketitle
\newcommand{\tDelta}{\widetilde{\Delta}}
\newcommand{\tmu}{\widetilde{\mu}}
\newcommand{\tE}{\widetilde{E}}
\newcommand{\tu}{\widetilde{u}}
\newcommand{\tv}{\widetilde{v}}
\newcommand{\sgn}{\operatorname{sgn}}

Systems of ultracold atoms provide one of the most versatile platforms for realizing many-body 
quantum phases of matter. In fact, several quantum phases and phase transitions 
such as the superfluid-Mott transition~\cite{Jaksch,Greiner,Folling,Spielman,Campbell,Bakr} have been realized 
in such systems. Yet, many of the most interesting phases or phase transitions in such systems are yet to 
be observed. One of the most glaring examples is the elusive antiferromagnetic N\'{e}el order~\cite{Greif,Hart}
 in the fermionic Hubbard model, 
which is believed to be a precursor of superconductivity in the model. Another example is the recently proposed family of phases based on the realization of spin-orbit coupling (SOC) by artificial gauge fields~\cite{lin,
wang,cheuk, Zhang1, Qu}, which includes 
topological insulators~\cite{Beri,Spielman12,Lewenstein}, topological superfluids (TSFs)~\cite{Liang,Liuxj,Wei,Zhang,Sato,jay11}, and fractional quantum Hall 
phases~\cite{Cooper}. A generic obstruction to the observations of many of these phases is heating due to 
spontaneous emission from applied laser fields. The heating problem makes it difficult to cool into the equilibrium thermal 
state of many of these topological phases. To study these phases, one can also prepare a gapped nontopological state 
and ramp the Hamiltonian to drive the system from the nontopological to the topological state. However, the properties of the short-lived topological phase are difficult to probe while it is subject to thermal fluctuations.

In this Letter, we propose a dynamical solution to the problem of studying the short-lived topological phase  by 
starting the system in
 its long-lived nontopological phase and driving it into the topological phase and back. 
 The rapid
nature of this process obviates heating; this is expected to make our proposal easily
implementable in experiments. The process involves crossing the 
quantum phase transition between the phases, which supports gapless excitations. Driving through the 
gapless phase transition produces excitations in the gapped phase via the Landau-Zener (LZ) 
transitions~\cite{landau,zener} with a defect density that demonstrates Kibble-Zurek (KZ) scaling~\cite{Kibble,Zurek,Sadler,Weiler,Lamporesi,Chen,Navon, Braun,Corman,Chomaz}. More interestingly, our dip-in-dip-out 
strategy, where the system is driven through the phase transition and back, leads to the St\"{u}ckelberg
interference phenomenon~\cite{Stuckelberg,nori} between the two LZ transitions, which in turn results in oscillations of the momentum and energy 
distribution of the excitations with the ramp rate. In many cases the unique ramp-rate dependence of the excitations' momentum distributions can be measured via standard time-of-flight techniques. This provides an experimentally 
viable test for the dynamical fingerprints of the topological phase transition (TPT), whose equilibrium properties would otherwise be hard to access.

While this general idea applies to many phase transitions in ultracold bosonic and fermionic systems~\cite{Greiner02,Sengupta12,Sengupta14,Braun}, we focus on phase transitions 
whose dynamical properties are well understood~\cite{Zhang,Sato,jay11,Bermudez09,Bermudez10,DeGottardi11,Perfetto13,Rajak14,Sengupta14,Kells14,Sacramento14,Hegde15}. In particular, we apply this idea to the proposed TSFs~\cite{Zhang,Sato,jay11} in systems of ultracold atoms which host the Majorana modes~\cite{Alicea12,Leijnse,Beenakker,Tudor,Franz}.
 Two of
the key ingredients~\cite{jay2} for realization of TSFs, namely,
controllable Zeeman coupling
 and fermionic Cooper pairing are readily available
in cold atomic systems. The recent realization of synthetic
SOC in cold atoms~\cite{lin,
wang,cheuk, Zhang1, Qu}
provides the third critical ingredient for
realizing topological superfluidity thus opening up the
possibility of observing topological phases in ultracold atomic
setting. In addition, the challenges of spatial and energy-resolved spectroscopy are easily resolved~\cite{Wei,Sylvain}. Despite the advantages of these proposals, the detection of TSFs in cold atomic systems is made
difficult by the low temperature scales involved combined with the heating
associated with SOC.

For the one-dimensional (1D) spin-orbit-coupled Fermi gases (SOCFGs) studied here, the TPT is accessed by raising the 
Zeeman field past a critical value~\cite{Liang,Liuxj,Wei,jay2}. Using the self-consistent time-dependent Bogoliubov--de Gennes equation (td-BdGE) 
formalism, we calculate the 
spin-resolved momentum distribution (SRMD) of the SOCFGs as it is ramped across the TPT through our dip-in-dip-out protocol described earlier. We find that the dynamics of the SRMD reflect both
 St\"{u}ckelberg
interference phenomenon and KZ scaling behavior for appropriate experimentally accessible ramp rates. 
We demonstrate that these oscillations and the scaling behavior persist at finite initial temperature and are
robust features of the TPT separating the conventional and topological phases of the Fermi
superfluids (SFs). While a gap closing is not by itself unique to TSFs, 
a closing of the gap of the nondegenerate Bogoliubov quasiparticles spectrum at zero momentum~\cite{kitaev} is 
 a yet experimentally unobserved smoking-gun signature for a TPT.

\begin{figure}[h]
\capstart
\begin{center}
\includegraphics{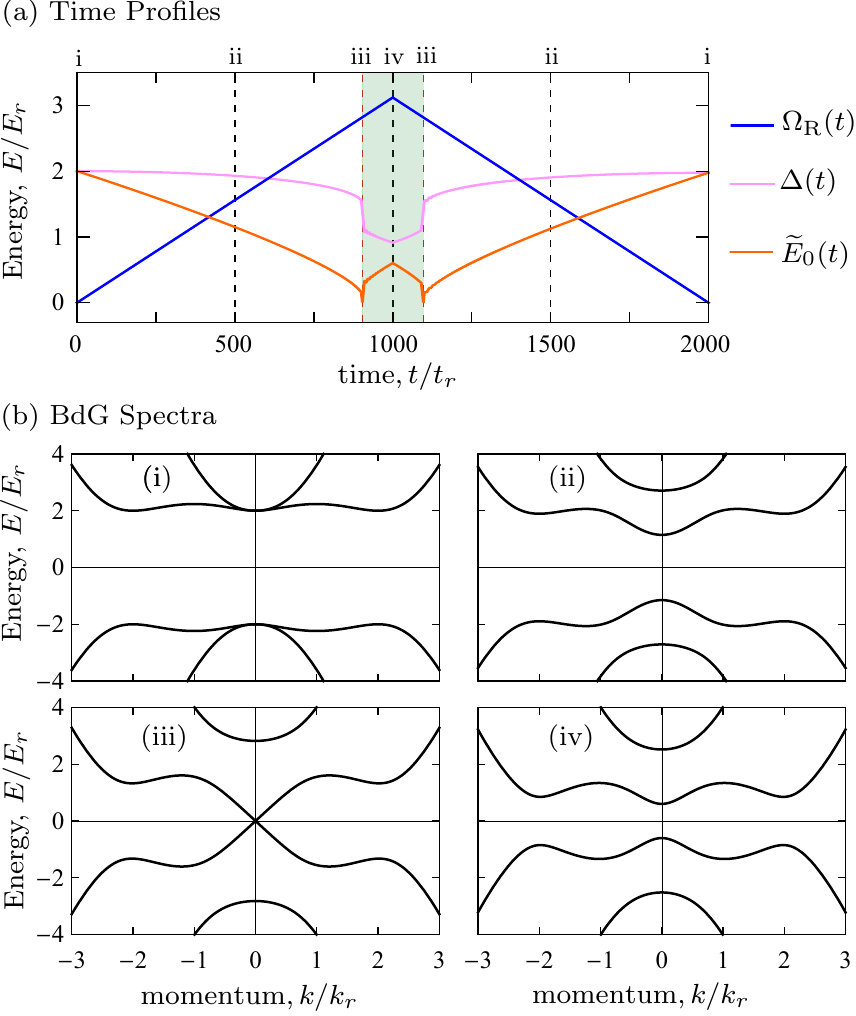}
\end{center}
\colorcaption{(a) Time profiles of 
$\Omega_{\mathrm{R}}(t)$, $\Delta(t)$,  and $\tE_0(t)$ for
$t_{\mathrm{ramp}} = 1000 t_r$. The dashed lines denote the times whose instantaneous band diagrams are plotted in (b). The red dashed lines mark the
critical times when TPT happens, and the shaded region corresponds to
the topological regime. Plots are obtained from numerically solving the td-BdGE [Eq.~\eqref{eq:TDBDG}] self-consistently [Eqs.~\eqref{eq:selfconst} and ~\eqref{eq:selfconst1}] with initial parameters: $\Omega_{\mathrm{R}}(0) =0$, $\Delta(0) = 2E_r$ and $\mu(0) = 0$ for SOC strength $\alpha = 2 E_r/k_r$ and
$t_{\mathrm{ramp}} = 1000t_r$. (b) Quasiparticle spectra at different
Zeeman potentials $\Omega_{\mathrm{R}}$. From top to bottom, the energy bands are labeled by $E_{2,k},E_{1,k},E_{-1,k}$, and $E_{-2,k}$. The parameters are as follows: (i) $\Omega_{\mathrm{R}} = 0$, $\Delta = 2
E_r$, $\mu = 0$, (ii) $\Omega_{\mathrm{R}} = 1.56E_r$, $\Delta = 1.93E_r$, $\mu
= -0.02E_r$, (iii) $\Omega_{\mathrm{R}} = 2.8 E_r$, $\Delta = 1.4E_r$, $\mu
= -0.14E_r$, and (iv) $\Omega_{\mathrm{R}} = 3.12E_r$, $\Delta = 0.91E_r$,
$\mu = -0.3E_r$.} \label{fig:bandstructure}
\end{figure}

We study 1D fermionic atoms with SOC and attractive $s$-wave interactions. The SOC is generated by a pair of counterpropagating Raman lasers, with recoil wave vector $k_r$, energy $E_r = \hbar^2k_r^2/2m$,
and characteristic time scale $t_r= \hbar/E_r$, giving the SOC strength $\alpha =
\hbar^2 k_r/m$.  These lasers couple two hyperfine atomic states representing the pseudospins $\sigma=
\uparrow, \downarrow$ (for example, $\left|\uparrow\right\rangle
\equiv \left|f = 9/2,m_F = -7/2\right\rangle$ and
$\left|\downarrow\right\rangle \equiv \left|f = 9/2,m_F =
-9/2\right\rangle$ in $^{40}$K atoms~\cite{Williams13}). The transverse
Zeeman potential strength $\Omega_{\mathrm{R}}$, set by the Raman coupling strength~\cite{lin}, is varied in time to drive the TPT. Here we consider
varying $\Omega_{\mathrm{R}}$ linearly from 0 to $\Omega_{\mathrm{R}f}$ in a time $t_{\mathrm{ramp}}$, and back in the same time: a piecewise linear ramp protocol of duration $2t_{\mathrm{ramp}}$ [see blue curve in Fig.~\ref{fig:bandstructure}(a)]. Because our protocol starts with Raman lasers off ($\Omega_{\mathrm{R}} = 0 $), it is straightforward to experimentally realize a long-lived conventional SF as the initial state~\cite{Deborah}; as we will see below, $t_{\mathrm{ramp}}$ is much less than the system's lifetime (either limited by the spontaneous emission of the Raman lasers or inelastic scattering from the Feshbach resonances). 

The
system's Hamiltonian in the Nambu basis $\Psi_k(t) = (\psi_{
k\uparrow}(t),\psi_{k\downarrow }(t),\psi^\dagger_{-k\downarrow}(t),
-\psi^\dagger_{-k\uparrow}(t))^\top$ is $H(t) = \frac{1}{2}\int dk
\Psi_k(t)^\dagger \mathcal{H}_{\mathrm{BdG},k}(t) \Psi_k(t)$, where
$\psi_{k \sigma}(\psi^\dagger_{k \sigma})$ denote the annihilation (creation) operators for fermions
with momentum $k$ and spin $\sigma$. The Bogoliubov--de Gennes (BdG) Hamiltonian is
\cite{roman,oreg,Wei,sau1}
\begin{equation}\label{eq:BdG}
\mathcal{H}_{\mathrm{BdG},k}(t) = \xi_k(t)\tau_z +
\alpha k\tau_z\sigma_z + \frac{\Omega_{\mathrm{R}}(t)}{2}\sigma_x + \Delta(t)\tau_x,
\end{equation}
where $\boldsymbol\sigma$ and $\boldsymbol\tau$ are vectors of Pauli operators acting on
spin and particle-hole space, respectively. Here, $\xi_k(t) = \hbar^2k^2/2m - \mu(t)$
combines the kinetic energy and the chemical potential $\mu(t)$, which is determined self-consistently to keep the
number of atoms fixed.

The mean-field
pairing potential 
\begin{equation}\label{eq:interact}
\Delta(t)e^{i\vartheta(t)}  = g_{\mathrm{1D}}\int{\langle\psi_{k\uparrow}(t)\psi_{
-k\downarrow}(t)\rangle dk}
\end{equation}
is also self-consistently determined,
where $\langle\dots\rangle$ denotes averaging with respect to the
initial thermal distribution. The attractive effective 1D coupling constant $g_{\mathrm{1D}}<0$ can be controlled by Feshbach tuning the three-dimensional (3D) scattering
length~\cite{bergeman,astra,liu}. In Eq.~\eqref{eq:BdG}, we used the transformed basis where $\psi_{k\sigma}(t)
\rightarrow \psi_{k\sigma}(t)\exp[i\vartheta(t)/2] $, giving a real pairing potential: $\Delta(t)\exp[i\vartheta(t)] \rightarrow
\Delta(t)$.

The instantaneous quasiparticle excitation spectrum of the BdG
Hamiltonian [cf.~Fig.~\ref{fig:bandstructure}(b)] consists of four
bands, $E_{n,k} = \sgn(n)\epsilon_{(-1)^{n},k}$, where $n = \pm 1,
\pm 2$ and
\begin{eqnarray}
\epsilon^2_{\pm,k}(t) &=& \frac{\Omega_{\mathrm{R}}(t)^2}{4} + \Delta(t)^2 + \xi_k(t)^2 +
\alpha^2k^2  \label{espec} \\
& &\pm 2\sqrt{\xi_k(t)^2\left[\alpha^2k^2 + \frac{\Omega_{\mathrm{R}}(t)^2}{4}\right]+\Delta(t)^2\frac{\Omega_{\mathrm{R}}(t)^2}{4}}.
\nonumber\
\end{eqnarray}
Since  $\mathcal{H}_{\mathrm{BdG},k}$ respects particle-hole
symmetry, the spectrum is symmetric around $E = 0$. As shown in Fig.~\ref{fig:bandstructure}(b), the instantaneous energy spectrum is gapped for $k\ne0$; however, for $k = 0$ the gap closes when
$\epsilon_{-,0}(t) = \Omega_{\mathrm{R}}(t)/2 -\sqrt{\Delta(t)^2 + \mu(t)^2}=0$.
Such a gap closing without change in the symmetry of the ground
state (which remains SF for all $\Omega_{\mathrm{R}}$) signifies a TPT~\cite{roman,jay2,oreg}
between topological $[\epsilon_{-,0}(t)
>0]$ and conventional SF phases $[\epsilon_{-,0}(t)<
0]$. For $\Omega_{\mathrm{R}} = 0$, the positive
and negative bands are doubly degenerate at $k = 0$; any nonzero $\Omega_{\mathrm{R}}$
lifts this degeneracy. 

To study the dynamics around the TPT, we propose to prepare conventional SFs
 $[\epsilon_{-,0}(t)<
0]$ at nonzero temperature $T$. We then drive the system through the TPT by
changing $\Omega_{\mathrm{R}}$ according to our ramp protocol with $\Omega_{\mathrm{R}f} >
2\sqrt{\Delta_f^2 + \mu_f^2}$ (where the subscript $f$ denotes the quantities at time $t = t_{\mathrm{ramp}}$) such that the ramp crosses the TPT (cf. Fig.~\ref{fig:bandstructure}).

We first analytically study the dynamics, considering the simple case of slow ramps at $T=0$. In this
limit, excitations occur near $k = 0$ and at the transition
times $t=t_{c(1,2)}$, given by the roots of
$\Omega_{\mathrm{R}}(t_c) = 2\sqrt{\Delta(t_{c})^2 +\mu(t_c)^2}$, where the Fermi gas changes from conventional to
TSF and vice versa. For 
$\hbar^2k^2/2m \ll \alpha k$, we approximate
\begin{equation}
\mathcal{H}_{\mathrm{BdG},k}(t)
\approx \alpha k \tau_z\sigma_z - \mu(t)\tau_z +
\frac{\Omega_{\mathrm{R}}(t)}{2}\sigma_x +\Delta(t)\tau_x.
\end{equation}
In this limit, excitations occur only between the $E_{1,k}$ and
$E_{-1,k}$ bands [cf. Fig.~\ref{fig:bandstructure}(b)]. At $k = 0$, the
eigenenergies are $ \pm\tE_0(t)$, where $\tE_0(t) =  |
\sqrt{\Delta(t)^2+\mu(t)^2}-\Omega_{\mathrm{R}}(t)/2|$ with eigenstates
\begin{subequations}\label{eq:phi0pm}
\begin{align}
\widetilde{\phi}_0^+(t) &=
\left(
\renewcommand{\arraystretch}{1.3}
\begin{matrix}
\cos \frac{\theta(t)}{2}\\ \sin \frac{\theta(t)}{2}
\end{matrix} \right)\otimes 
\frac{1}{\sqrt{2}}
\left( 
\renewcommand{\arraystretch}{1.3}
\begin{matrix}
1 \\ 1
\end{matrix}
\right),\\
\widetilde{\phi}_0^-(t) &=
\left(
\renewcommand{\arraystretch}{1.3}
\begin{matrix}
-\sin \frac{\theta(t)}{2} \\ \cos \frac{\theta(t)}{2}
\end{matrix}
\right)
\otimes 
\frac{1}{\sqrt{2}}
\left(
\renewcommand{\arraystretch}{1.3}
\begin{matrix}
 1 \\ -1
\end{matrix}
\right),
\end{align}
\end{subequations}
where $\widetilde{\phi}_0^\pm(t)$ corresponds to positive
and negative bands [with pseudospin $\left|\pm\right\rangle \equiv
(\left|\uparrow\right\rangle \pm
\left|\downarrow\right\rangle)/\sqrt{2}$] and $\cos\theta(t) \equiv
\mu(t)/\sqrt{\Delta(t)^2 + \mu(t)^2}$. In
the subspace of these eigenstates, the effective low-energy Hamiltonian
near $k = 0$ is
\begin{eqnarray}\label{eq:HBdGp}
\widetilde{\mathcal{H}}_{\mathrm{BdG},k}(t) &=&
\widetilde{\alpha}(t)k\eta_x + \tE_0(t)\eta_z, \label{effham}
\end{eqnarray}
where $\widetilde{\alpha}(t) = \alpha \sin \theta(t)$, $\eta_x =
\widetilde{\phi}_0^+(t)[\widetilde{\phi}_0^-(t)]^\dagger +
\mathrm{H.c.}$, $\eta_z =
\widetilde{\phi}_0^+(t)[\widetilde{\phi}_0^+(t)]^\dagger -
\widetilde{\phi}_0^-(t)[\widetilde{\phi}_0^-(t)]^\dagger$, and
$2\eta_y = -i[\eta_z,\eta_x]$. Equation~\eqref{effham}
is a two-parameter driven Hamiltonian~\cite{sau1} with
instantaneous energy eigenvalues $ \pm\tE_k(t)$, where $\tE_k(t) =
\sqrt{\tE_0(t)^2 + \widetilde{\alpha}(t)^2 k^2}$.

We analyze the dynamics of the TPT using
$\widetilde{\mathcal{H}}_{\mathrm{BdG},k}(t)$, where the
single-particle state of the system at time $t$ is given by
\begin{equation}\label{eq:spsstate}
\widetilde{\phi}_k(t) = b_k^+(t)\left(\begin{matrix} w^+_k(t)\\
\sgn(k)w^-_k(t) \end{matrix}\right) +
b_k^-(t)\left(\begin{matrix}- \sgn(k)w^-_k(t)\\
w^+_k(t) \end{matrix}\right),
\end{equation}
with the initial conditions $b_k^+(0) = 0 $ and $b_k^-(0) = 1$. These
two-component vectors are expressed in the basis
$\widetilde{\phi}_0^\pm$ with $w^{\pm}_k(t) = \sqrt{[
1 \pm \tE_0(t)/\tE_k(t)]/2}$. The Schr\"{o}dinger equation for the
system then leads to 
\begin{equation}\label{eq:schro}
i\hbar\partial_t\vec{b}_k(t) =
\widetilde{\mathcal{H}}_{\mathrm{BdG},k}(t) \vec{b}_k(t),
\end{equation}
where
$\vec{b}_k(t) = (b_k^+(t),b_k^-(t))^\top$.

\begin{figure}[h!]
\capstart
\begin{center}
\includegraphics{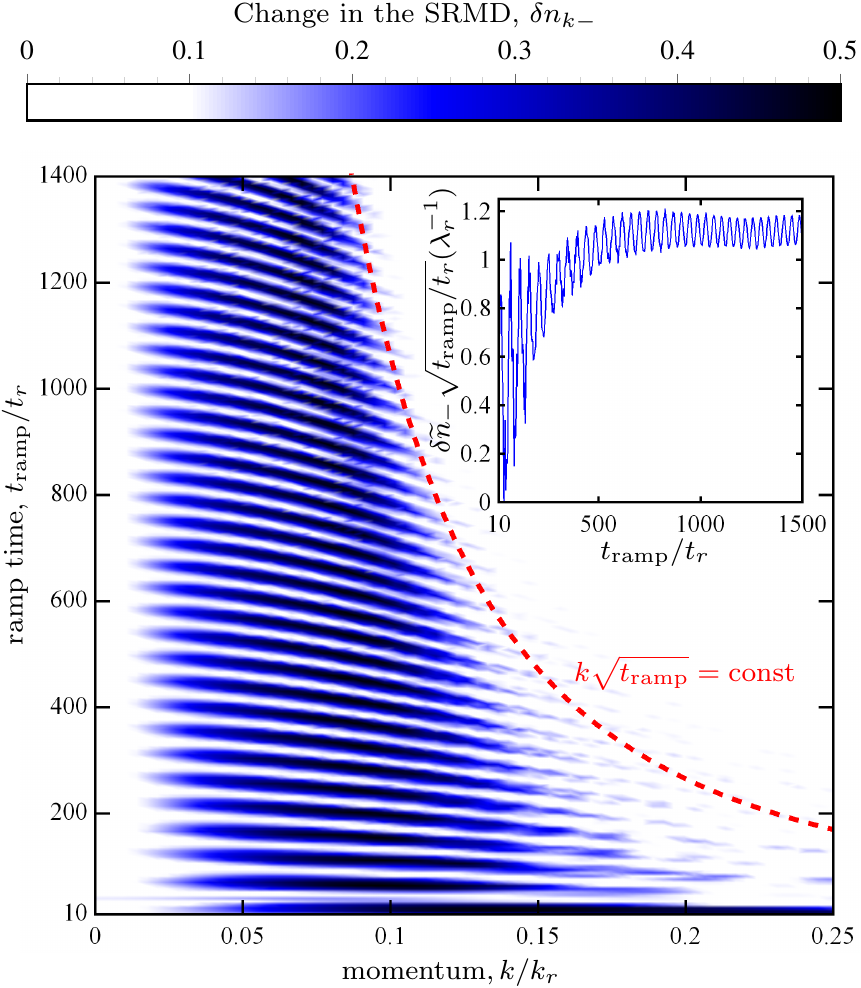}
\end{center}
\colorcaption{Change in the SRMD $\delta n_{k -}$ for spin
$\left|-\right\rangle = \left(\left|\uparrow\right\rangle -
\left|\downarrow\right\rangle\right)/\sqrt{2}$ as a function of
$t_{\rm ramp}/t_r$ and $k/k_r$. For large $t_{\rm ramp}$, the width of the oscillation envelopes scales with $1/\sqrt{t_{\mathrm{ramp}}}$ as shown by the red dashed line. $\delta n_{k-}$ is symmetric with respect to $k =0$; thus, for illustration purposes, we only plot $\delta n_{k-}$ for $k \geq 0$. Note that $\delta n_{k+} = - \delta n_{k -}$. Inset: Integrated change in SRMD
$\delta\widetilde{n}_- = \int{dk \delta n_{k-}}$ as a function of $t_{\rm ramp}/t_r$
exhibiting oscillations, with the amplitude of the
oscillations at large $t_{\rm ramp}$ scaling like $\sqrt{t_{\rm
ramp}}$, as can be read off directly from the $y$ axis. The plots are obtained by numerically solving
Eq.~\eqref{eq:TDBDG} self-consistently [Eqs.~\eqref{eq:selfconst} and \eqref{eq:selfconst1}] with initial
conditions $\mu(0) = 0$, $\Delta(0) = 2 E_r$, and $\Omega_{\mathrm{R}} (0) = 0$ for a
temperature $k_{\mathrm{B}}T = 0.1E_\mathrm{F}$ (which is below the critical temperature $T_c = 0.19T_\mathrm{F}$~\cite{Nascimbene,Zwierlein}), SOC strength $\alpha = 2 E_r/k_r$, and $\Omega_{\mathrm{R}f}=3.12 E_r$. \label{fig:excitmomen}}
\end{figure}

We make further analytical progress by ignoring the self-consistency
condition so that the system can be treated as a collection of
two-level systems for each $(k,-k)$ pair and use the
adiabatic-impulse approximation
\cite{damski05,damski06,nori,adutt1,Dziarmaga, note} that describes
such periodic dynamics accurately for low frequency and/or large
amplitude drives. Within this approximation, excitations are produced only near the critical gap-closing
times $t_{c(1,2)}$ when the system enters the impulse regime; otherwise, the dynamics occur adiabatically in each band and the system
accumulates a dynamical phase $U(t_f,t_i) = \exp[-i\eta_z
\int_{t_i}^{t_f}{dt \tE_k(t)/\hbar}]$. In the former regime, near
the gap-closing times $t_{c(1,2)}$, excitations are produced and the
evolution operator is~\cite{nori}
\begin{equation}
N = \sqrt{1-p_k}[i\sin(\varphi_{\mathrm{S},k}) - \eta_z\cos(\varphi_{\mathrm{S},k})]-i\eta_y \sqrt{p_k},
\end{equation}
where  $p_k = \exp\left({-2\pi\delta_k}\right)$ is
the probability of excitation formation in each passage through the
critical point~\cite{landau,zener} with $\delta_k =
(\alpha k)^2/(2\hbar|d\tE_0(t)/dt|_{t_{c}})$, and $\varphi_{\mathrm{S},k} = \pi/4
+ \delta_k (\ln\delta_k - 1) + \arg\Gamma(1-i\delta_k)$ is the
Stokes phase originating from the interference of the parts of the
system wave function in the instantaneous ground and excited states
at $t=t_{c(1,2)}$ with $\arg\Gamma(1-i\delta_k)$ being the argument of the gamma function~\cite{gradshteyn}. These results give the probability of defect formation 
\begin{equation}
P^{\mathrm{ex}}_k = 4 p_k(1-p_k)\sin^2 \Phi_{\mathrm{St},k}
\end{equation}
 at $t=2t_{\mathrm{ramp}}$, where $\Phi_{\mathrm{St},k}= \zeta_{2k} + \varphi_{\mathrm{S},k}$ is
the St\"{u}ckelberg phase and $\zeta_{2k} = \int_{t_{c1}}^{t_{c2}}dt
\tE_k(t)/\hbar$ is the dynamical phase factor accumulated during
passage between the two crossings of the gap-closing points
\cite{nori,adutt1,note}. Since the excitations occur near $k
\sim 0$ where the $E_{\pm1,k}$ band approximately corresponds to pseudospin
$|\pm\rangle$ (along the $x$ direction), $P^{\mathrm{ex}}_k$ is directly related to changes
in the SRMD $\delta n_{k\pm}$ measured along the pseudospin $x$ direction. Furthermore, within these
approximations, $|d\widetilde{E}_0(t)/dt|_{t_{c(1,2)}} =
\Omega_{\mathrm{R}f}/(2t_\mathrm{ramp})$, and it can be shown that $P_k^{\mathrm{ex}}$ is a
function of $k\sqrt{t_{\mathrm{ramp}}}$ only (see Ref.~\cite{note} for the derivation). Thus, the integrated change of the SRMD $\delta {\widetilde n}_\pm = \int dk
\delta n_{k\pm}$ displays KZ scaling $\sim \sqrt{t_{\rm ramp}}$ of defect density for a system dynamically evolved through the TPT. We now show that these properties persist even when the
self-consistency conditions for $\Delta(t)$ and $\mu(t)$ are imposed, as well as at nonzero $T$ (see 
Fig.~\ref{fig:excitmomen}).

We solve for the dynamics of the single-particle density matrix $\rho^{ab}_k(t) = \langle \Psi^{\dagger a}_k(t) \Psi^{b}_k(t)
\rangle$  self-consistently and at finite initial temperature,
where $a,b$ denote the indices of elements in the Nambu basis. The density matrix obeys the
equation of motion [Eq.~\eqref{eq:BdG}]
\begin{equation}\label{eq:TDBDG}
i\hbar\partial_t \rho_k (t) =
[\mathcal{H}_{\mathrm{BdG},k}(t),\rho_k(t)],
\end{equation}
subject to the self-consistency conditions (see Ref.~\cite{note}
for the derivation)
\begin{subequations}
\begin{align}
\Delta(t)&= \frac{g_{\mathrm{1D}}}{4}\int {dk  \mathrm{Tr}(\rho_k(t) \tau_x)},\label{eq:selfconst}\\
\mu(t) &=  \frac{g_{\mathrm{1D}}}{4\Delta(t)} \int dk \mathrm{Tr}\left( \rho_k(t) \Lambda_k(t)
\right),\label{eq:selfconst1}
\end{align}
\end{subequations}
where  $\Lambda_k(t) =  \left(\hbar^2k^2/2m + \alpha k \sigma_z\right)\tau_x
-\Delta(t)\tau_z$. Our system begins in the thermal state
\begin{equation}
\rho_{k}(t)  =  \sum_{\substack{n\\
E_{n,k}(0)<0}} f_{n,k}
\chi_{n,k}(t)\chi_{n,k}^{\dagger}(t)
+(1-f_{n,k})\widetilde{\chi}_{n,-k}(t)
\widetilde{\chi}_{n,-k}^{\dagger}(t), \label{eq:init}
\end{equation}
where $f_{n,k}= [\exp(E_{n,k}(0)/k_{\mathrm{B}}T)+1]^{-1}$ is the Fermi function of
the initial Hamiltonian, and $k_{\mathrm{B}}$ is Boltzmann's constant. The
wave function $\chi_{n,k}(t)$ with its particle-hole conjugate
$\widetilde{\chi}_{n,k}(t) = \tau_y \sigma_y \chi_{-n,-k}^{\ast}(t)$
begins as eigenfunctions of the initial
Hamiltonian and evolves according to $i\hbar\partial_t\chi_{n,k}(t) =
\mathcal{H}_{\mathrm{BdG},k}(t)\chi_{n,k}(t)$. 
Figure~\ref{fig:bandstructure}(a) shows the resulting time profiles of the  pairing potential obtained from solving the td-BdGE (see Ref.~\cite{note} for the time dependence of all parameters and remarks on the numerical simulation).

We numerically solved the td-BdGE
for the change in the SRMD 
\begin{equation}
\delta n_{k\pm} =
\mathrm{Tr}\left(\left[\rho_k(2t_\mathrm{ramp}) - \rho_k(0) \right]
\left[\left(\frac{1+\tau_z}{2}\right)\otimes\left(\frac{1\pm\sigma_x}{2}\right)\right]\right).
\end{equation}
Figure~\ref{fig:excitmomen} shows that $\delta n_{k-}$
still exhibits St\"{u}ckelberg oscillations even with inclusion of the
self-consistency conditions and at $T > 0$. Furthermore, for $t_{\mathrm{ramp}}\gg \hbar/\Delta_f$, we still see $\delta
{n}_{k\pm} \sim k \sqrt{t_{\rm ramp}}$ (see Ref.~\cite{note} for an explicit demonstration of the scaling), and the integrated change in SRMD $\delta {\widetilde n}_\pm = \int dk
\delta {n}_{k\pm}$ therefore scales with
$\sqrt{t_{\mathrm{ramp}}}$, thus, showing the robustness of such interference phenomenon
in the present system. We verified that these features appear
only if $\Omega_{\mathrm{R}f} > 2\sqrt{\Delta_f^2 + \mu_f^2}$, where the ramp
takes the system through the TPT; thus both the KZ scaling and the
presence of St\"{u}ckelberg oscillations mark the TPT. In our
calculation, we ignored the effect of phase fluctuation as this effect can be suppressed by coupling an array of 1D SOCFGs~\cite{mizushima,sau2,meng,fidkowski}.

The parameters
used for the plots in Fig.~\ref{fig:excitmomen} are realistic for 1D SOCFG
experiments. For experiments with $^{40}$K, the Raman laser beams,
coupling the $\left|\uparrow\right\rangle \equiv
\left|9/2,-7/2\right\rangle$ and $\left|\downarrow\right\rangle
\equiv \left|9/2,-9/2\right\rangle$ states, have laser wavelength
$\lambda_r = 768.86$ nm, giving the recoil energy $E_r = h \times 8.445$ kHz,
and time $t_r = \hbar/ E_r \approx 20$ $\mu$s~\cite{Williams13}. The
single-body decay time due to photons scattering from the Raman
lasers is about 60 ms~\cite{Williams13}, and the lifetime
owing to three-body recombination is about 200 ms~\cite{Jin04}. We consider SOCFGs
with Fermi energy $E_{\mathrm{F}} = E_r$. The 1D
Fermi gas criterion is satisfied when $E_{\mathrm{F}} < \hbar\omega_\perp$; for the
lateral trapping frequency $\omega_\perp/2\pi =  5
\times 10^4$ Hz,  which corresponds to characteristic harmonic
oscillator length $d_\perp =\sqrt{\hbar/m\omega_\perp} \approx 1345a_0$, where $a_0$ is the Bohr radius; the parameters
used in the calculation for the plots in Fig.~\ref{fig:excitmomen}
correspond to linear density $\widetilde{n} \approx 5$ $\mu \mathrm{m}^{-1}$ and 1D interaction strength $g_{\mathrm{1D}} \approx -0.73 E_r\lambda_r$ (or
3D scattering length $a_{\mathrm{3D}} \approx -2870a_0$~\cite{liu}).
For these values, Fig.~\ref{fig:excitmomen} shows that
the St\"{u}ckelberg oscillations and KZ scaling behavior of
the SRMD can be observed within the
experimentally limiting single-body decay time $(\approx 3000t_r)$ and thus
is feasible experimentally.

Our dip-in-dip-out protocol is quite general and can be gainfully used 
for observing features related to quantum phase transitions between long-lived and short-lived phases 
of ultracold bosonic and fermionic atoms. In addition, it provides a route to escaping the heating problem, which is one of the major obstacles in measuring properties of such systems in or near their short-lived phases.
Moreover, our work also shows that such a protocol applied to ultracold atom systems, including the one we analyzed in detail, may provide us with test beds for observation of both KZ scaling~\cite{Sadler,Weiler,Lamporesi,Chen,Navon, Braun,Corman,Chomaz} and St\"{u}ckelberg interference phenomenon~\cite{Mark, Kling,Zenesini}.

We thank H.-Y.~Hui, S.~S.~Natu, and J.~Radi\'{c} for useful discussions. F.~S. and J.~D.~S. acknowledge the support from LPS-CMTC, JQI-NSF-PFC and University of Maryland startup grants. I.~B.~S. gratefully acknowledges funding from the ARO's Atomtronics-MURI, the AFOSR's quantum matter MURI, the NSF through the JQI Physics Frontier Center, and NIST.

\onecolumngrid
\vspace{1cm}
\begin{center}
{\bf\large Supplemental Material for ``Dynamical Detection of Topological Phase Transitions in Short-Lived Atomic Systems"}
\end{center}
\vspace{0.5cm}

\setcounter{secnumdepth}{3}
\setcounter{equation}{0}
\setcounter{figure}{0}
\renewcommand{\theequation}{S-\arabic{equation}}
\renewcommand{\thefigure}{S\arabic{figure}}
\renewcommand\figurename{Supplementary Figure}
\renewcommand\tablename{Supplementary Table}
\newcommand\Scite[1]{[S\citealp{#1}]}

\makeatletter \renewcommand\@biblabel[1]{[S#1]} \makeatother

\section{Adiabatic-Impulse Approximation}
 The equation of motion $i\hbar\partial_t\vec{b}_k(t) =
\widetilde{\mathcal{H}}_{\mathrm{BdG},k}(t) \vec{b}_k(t)$ [Eq.~\eqref{eq:schro}], where $\vec{b}_k(t) = (b_k^+(t),b_k^-(t))^\top$ [Eq.~\eqref{eq:spsstate}] and $\widetilde{\mathcal{H}}_{\mathrm{BdG},k}(t) = \widetilde{\alpha}(t)k\eta_x + \tE_0(t)\eta_z$ [Eq.~\eqref{eq:HBdGp}] with $\eta_x$ and $\eta_z$ being the Pauli matrices acting on the subspace $\widetilde{\phi}_0^{\pm}(t)$ [Eq.~\eqref{eq:phi0pm}], can be expressed in form of two-decoupled second-order differential equations as
\begin{align}\label{eq:bk1}
\left\{-\hbar^2\partial_t^2 - \tE_k(t)^2 + i\hbar \left[\mp \partial_t\tE_0(t) \pm \tE_0(t) \partial_t-\frac{\partial_t\widetilde{\alpha}(t)}{\widetilde{\alpha}(t)}\left[i\hbar\partial_t\mp\tE_0(t)\right]\right]\right\}
b_k^{\pm} &= 0.
\end{align}
Assuming no self-consistency, we can use the adiabatic-impulse approximation~\cite{nori} to write Eq.~\eqref{eq:bk1} as $\vec{b}_k(t) = V\vec{b}_k(0)$ where the total evolution operator $V$ is decomposed into adiabatic $U$ and impulse $N$ operators. The adiabatic (impulse) regime corresponds to the time duration far away from (near) the critical gap-closing time
$t_{c(1,2)}$. In matrix form we can write down $U$ as
\begin{equation}
U_j = \left(\begin{matrix} e^{-i\zeta_{jk}} & 0 \\ 0 &
e^{i\zeta_{jk}}\end{matrix} \right), \hspace{0.3 cm} j = 1,2,3,
\end{equation}
where the dynamical phases are given by $\zeta_{1k} = \int_{0}^{t_{c1}}{dt \tE_k(t)/\hbar}$, $\zeta_{2k} = \int_{t_{c1}}^{t_{c2}}{dt \tE_k(t)/\hbar}$, and $\zeta_{3k} =
\int_{t_{c2}}^{2t_{\mathrm{ramp}}}{dt \tE_k(t)/\hbar}$. The impulse operator $N$ can be written as~\cite{nori}
\begin{equation}
\hspace{0.3 cm}  \hspace{0.5 cm} N= \left(\begin{matrix} \sqrt{1 - p_k}e^{-i\widetilde{\varphi}_{\mathrm{S},k}} & -\sqrt{p_k} \\ \sqrt{p_k} & \sqrt{1 -p_k}e^{i\widetilde{\varphi}_{\mathrm{S},k}} \end{matrix}\right), 
\end{equation}
where $p_k = \exp(2\pi\delta_k)$ is the landau-Zener transition probability~\cite{landau,zener} at each critical time, $\delta_k =
(\alpha k)^2/(2\hbar|d\tE_0(t)/dt|_{t_{c}})$, $\widetilde{\varphi}_{\mathrm{S},k} = \varphi_{\mathrm{S},k} -\pi/2$ and $\varphi_{\mathrm{S},k} = \pi/4 + \delta_k (\ln\delta_k - 1) + \arg\Gamma(1-i\delta_k)$. The Stokes phase $\varphi_{\mathrm{S},k}$ increases monotonously from $0$ in the adiabatic limit ($\delta_k \rightarrow \infty$) to $\pi/4$ in the diabatic or fast driving limit ($\delta_k \rightarrow 0$), as seen from the asymptotic argument of the gamma function~\cite{gradshteyn}
\begin{align}
\mathrm{arg}~\Gamma (1-i\delta_k) \approx \begin{cases} C \delta_k, &
\delta_k \ll 1, \\ -\frac{\pi}{4} - \delta_k(\ln\delta_k -1), & \delta_k \gg
1,
\end{cases}
\end{align}
where $C \approx 0.58$ is the Euler constant. At the end of the ramp protocol, the total evolution operator becomes
\begin{align}
V = U_3NU_2NU_1 = \left(\begin{matrix} \beta_k & -\gamma_k^* \\ \gamma_k & \beta_k^*\end{matrix}\right),
\end{align}
with matrix elements
\begin{align}
\beta_k &= (1-p_k)e^{-i\zeta_{+k}} - p_ke^{-i\zeta_{-k}}, \nonumber\\
\gamma_k &= \sqrt{(1-p_k)p_k} e^{i(\widetilde{\varphi}_{\mathrm{S},k}+2\zeta_{3k})}\left(e^{-i\zeta_{+k}} +e^{-i\zeta_{-k}}\right),
\end{align}
where the phases are given by $\zeta_{+k} = \zeta_{1k}+\zeta_{2k}+ \zeta_{3k}+  2\widetilde{\varphi}_{\mathrm{S},k}$ and $\zeta_{-k} = \zeta_{1k}- \zeta_{2k} +\zeta_{3k}$.
The probability of defect formation at the end of the ramp protocol (at $t = 2t_{\mathrm{ramp}}$) is then given by
\begin{equation}
P^{\mathrm{ex}}_k = |\gamma_k|^2= 4 p_k(1-p_k)\sin^2\Phi_{\mathrm{St},k},
\end{equation}
where $\Phi_{\mathrm{St},k} = \zeta_{2k} + \varphi_{\mathrm{S},k}$ is the St\"{u}ckelberg phase. Note that in the case of no-self consistency, $|d\widetilde{E}_0(t)/dt| = \Omega_{\mathrm{R}f}/(2t_{\mathrm{ramp}})$, and consequently $\delta_k$ is a function of $k \sqrt{t_{\mathrm{ramp}}}$. Since $p_k$ and $\varphi_{\mathrm{S},k}$ are functions of $\delta_k$, $P^{\mathrm{ex}}_k$ is also a function of $k\sqrt{t_{\mathrm{ramp}}}$. As a result, the defect density displays Kibble-Zurek scaling $\sim \sqrt{t_{\mathrm{ramp}}}$.

\section{Self-Consistency Condition}
The self-consistent chemical potential $\mu(t)$ [Eq.~\eqref{eq:selfconst1}] is derived from the constraint on the particle density $\widetilde{n}$, i.e.,
\begin{equation}\label{eq:density}
\int dk\mathrm{Tr}\left(\rho_k(t)\left(\frac{1+\tau_z}{2}\right)\right) = \widetilde{n}.
\end{equation}
Taking the time derivative of Eq.~\eqref{eq:density}, i.e., $i\hbar\partial_t \rho_k (t) =
[\mathcal{H}_{\mathrm{BdG},k}(t),\rho_k(t)]$, and using the cyclic property of trace, we have
\begin{align}\label{eq:firstderiv}
\frac{1}{2} \int{dk \mathrm{Tr} ([\mathcal{H}_{\mathrm{BdG},k}(t),\rho_k(t)]\tau_z)} &= 0\nonumber\\
\frac{1}{2} \int{dk \mathrm{Tr}([\tau_z,\mathcal{H}_{\mathrm{BdG},k}(t)]\rho_k(t))} &= 0\nonumber\\
\int{dk \mathrm{Tr}(\tau_y\rho_k(t))} &= 0.
\end{align}
Differentiating Eq.~\eqref{eq:firstderiv} with respect to time and using the cyclic property of trace, we then obtain
\begin{align}
\int{dk \mathrm{Tr}(\tau_y [\mathcal{H}_{\mathrm{BdG},k}(t),\rho_k(t)])} &= 0\nonumber\\
\int{dk \mathrm{Tr}([\mathcal{H}_{\mathrm{BdG},k}(t),\tau_y]\rho_k(t))}& = 0\nonumber\\
\int{dk \mathrm{Tr}\left((\hbar^2 k^2/2m-\mu(t) + \alpha k\sigma_z)\tau_x\rho_k(t) -\Delta(t)\tau_z\rho_k(t)\right)}& = 0.
\end{align}
Noting that $g_{\mathrm{1D}}\int{dk \mathrm{Tr}(\rho_k (t)\tau_x)}/4 = \Delta(t)$, we then have the self-consistent chemical potential $\mu(t)$ as
\begin{align}
\mu(t) =  \frac{g_{\mathrm{1D}}}{4\Delta(t)}\int{dk \mathrm{Tr}\left((\hbar^2 k^2/2m +\alpha k \sigma_z)\tau_x\rho_k(t)-\Delta(t)\tau_z\rho_k(t)\right)}.
\end{align}

\section{Remarks on The Numerical Simulation}

In the main text, the td-BdGE is given in terms of the single-particle density matrix $\rho_{k}(t)$. The td-BdGE can also be written in terms of the wave function $\chi_{n,k}(t) = (u_{n,k\uparrow}(t),u_{n,k\downarrow}(t), v_{n,k\downarrow}(t),-v_{n,k\uparrow}(t))^\top$ as  
\begin{equation}
i\hbar\partial_t\chi_{n,k}(t) = \mathcal{H}_{\mathrm{BdG},k}(t)\chi_{n,k}(t),
\end{equation}
subject to the self-consistency conditions
\begin{subequations}
\begin{align}
\Delta(t) &= \frac{g_{\mathrm{1D}}}{4} \sum_{\substack{n\\
E_{n,k}(0)<0}} \int dk \mathcal{I}^-_{n,k}(t), \\
\mu(t) &= \frac{g_{\mathrm{1D}}}{4 \Delta(t)} \sum_{\substack{n\\
E_{n,k}(0)<0}}\int dk \left[\frac{\hbar^2k^2}{2m}\mathcal{I}^-_{n,k}(t)+\alpha k \mathcal{I}^+_{n,k}(t)-\Delta(t)\mathcal{Q}_{n,k}(t)\right],
\end{align}
\end{subequations}
where
\begin{subequations}
\begin{align}
\mathcal{I}_{n,k}^{\pm}(t) &= (2 f_{n,k} -1)\{[v^*_{n,k\downarrow}(t)u_{n,k\uparrow}(t) \pm
u_{n,k\downarrow}(t)v^*_{n,k\uparrow}(t)] + \mathrm{H.c.}\},  \\
\hspace{0.3cm}\mathcal{Q}_{n,k}(t) &= (2 f_{n,k} -1)\sum_{\sigma}(|u_{n,k\sigma}|^2 -|v_{n,k\sigma}|^2),
\end{align}
\end{subequations}
with $f_{n,k}= [\exp(E_{n,k}(0)/k_{\mathrm{B}}T)+1]^{-1}$ being the Fermi function of the initial Hamiltonian.

\begin{figure}[h!]
\capstart
\begin{center}
\includegraphics{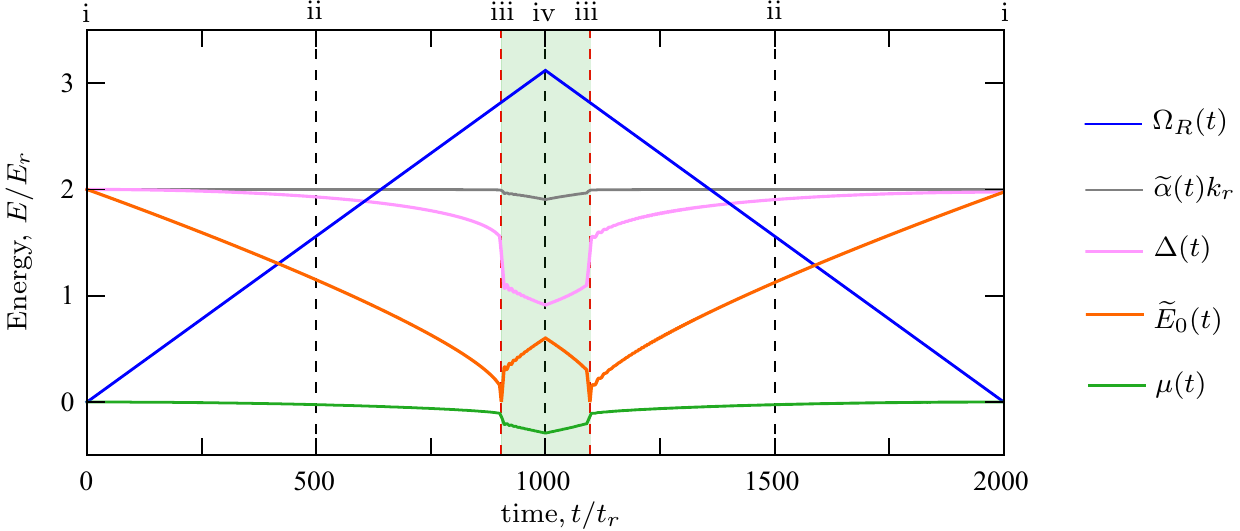}
\end{center}
\caption{Time profiles of $\Omega_{\mathrm{R}}(t)$, $\widetilde{\alpha}(t)$,
$\Delta(t)$, $\tE_0(t)$ and $\mu(t)$ for
$t_{\mathrm{ramp}} = 1000 t_r$. The dashed lines denote the times whose instantaneous band diagrams are plotted in Fig.~\ref{fig:bandstructure}(b) in the main text. The red dashed lines mark the
critical times when TPT happens, and the shaded region corresponds to
the topological regime. Plots are obtained from numerically solving the td-BdGE self-consistently with initial parameters: $\Omega_{\mathrm{R}}(0) =0$, $\Delta(0) = 2E_r$, and $\mu(0) = 0$ for SOC strength $\alpha = 2 E_r/k_r $ and
$t_{\mathrm{ramp}} = 1000t_r$.} \label{supplfig:timeprofiles}
\end{figure}

The self-consistent solution of the td-BdGE involves solving a large number of coupled time-dependent differential equations (one for each $k$ point). To reduce the number of time-dependent variables, we first calculated the self-consistent $\Delta(t)$ and $\mu(t)$ in the adiabatic regime by solving the time-independent BdG equation. The td-BdGE was then solved self-consistently for a small range of states near $k =0$ where excitations occur. Since the $\pm k$ eigenstates are related by $\mathrm{X}_{n,-k}=\sigma_x\mathrm{X}_{n,k}$, we accelerated the computation by focusing on $k\geq0$. Solving the td-BdGE self-consistently with the Zeeman potential $\Omega_{\mathrm{R}}(t)$ varied according the piecewise linear ramp protocol (see blue curve in Fig.~\ref{supplfig:timeprofiles}), we obtained $\widetilde{\alpha}(t)$,
$\Delta(t)$, $\tE_0(t)$, and $\mu(t)$ as shown in Fig.~\ref{supplfig:timeprofiles}.

\section{Spin-Resolved Momentum Distribution}

\begin{figure}[h!]
\capstart
\begin{center}
\includegraphics{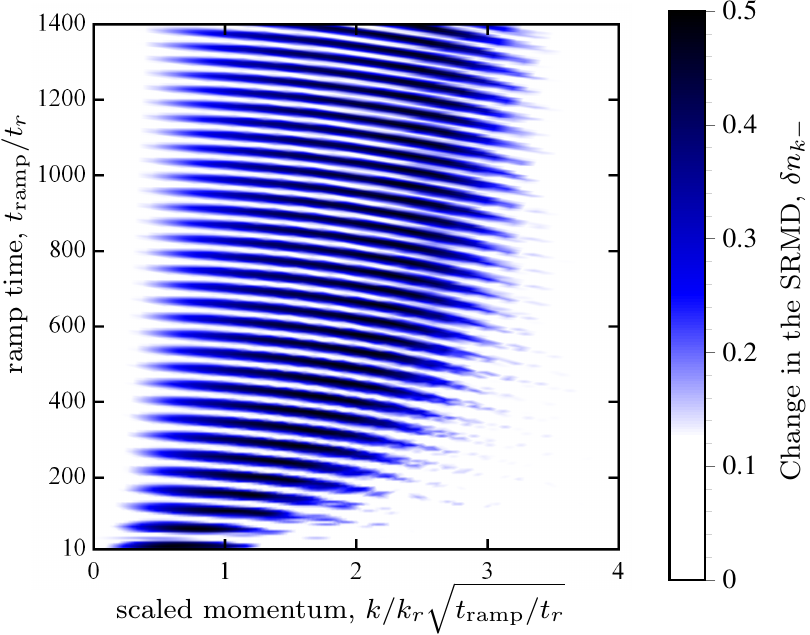}
\end{center}
\caption{Change in the SRMD $\delta n_{k -}$ for pseudospin
$\left|-\right\rangle = \left(\left|\uparrow\right\rangle -
\left|\downarrow\right\rangle\right)/\sqrt{2}$ as a function of
$t_{\rm ramp}/t_r$ and $k/k_r \sqrt{t_{\rm ramp}/t_r}$. Note that $\delta n_{k-}$ is a function of
$k\sqrt{t_{\rm ramp}}$ only for large $t_{\rm ramp}$, as seen from
its almost flat nature for small $k/k_r$ and the width of its
oscillation envelopes. The scaling of
$\delta n_{k -}$ can be read off directly from the $x$ axis. $\delta n_{k-}$ is symmetric with respect to $k =0$; thus, for illustration purposes, we only plot $\delta n_{k-}$ for $k \geq 0$. The plots are obtained by numerically solving the
td-BdGE self-consistently with initial
conditions $\Omega_{\mathrm{R}} (0) = 0$, $\Delta(0) = 2 E_r$ and $\mu(0) = 0$ for a
temperature $k_{\mathrm{B}}T = 0.1E_\mathrm{F}$ (which is below the critical temperature $T_c = 0.19T_\mathrm{F}$~\cite{Nascimbene, Zwierlein}), SOC strength $\alpha = 2 E_r/k_r$, and $\Omega_{\mathrm{R}f}=3.12 E_r$.  Note that $\delta n_{k+} = - \delta n_{k -}$ due to particle number conservation. }\label{supplfig:figs2}
\end{figure}

The change in spin-resolved momentum distribution $\delta n_{k-}$ shows St\"{u}ckelberg oscillations with the ramp time $t_{\mathrm{ramp}}$ and for large $t_{\mathrm{ramp}}$, $\delta n_{k-}$ scales with $\sqrt{t_{\mathrm{ramp}}}$, as shown in Fig.~\ref{fig:excitmomen} in the main text. In Fig.~\ref{supplfig:figs2}, we demonstrate the scaling more explicitly by plotting $\delta n_{k-}$ as a function of scaled momentum $k/k_r \sqrt{t_{\rm ramp}/t_r}$.

\end{document}